\documentclass[aps,prx,reprint,superscriptaddress,nofootinbib]{revtex4-2}
\usepackage{amsmath,amssymb,amsthm,microtype}
\usepackage[colorlinks=true,linkcolor=blue,citecolor=blue,urlcolor=blue]{hyperref}
\hypersetup{pdftitle={A spectral improvement of the Ashikhmin--Litsyn bound for quantum codes},pdfauthor={Yu-Xuan Zhang and Jing-Ling Chen}}
\newtheorem{theorem}{Theorem}
\DeclareMathOperator{\wt}{wt}
\DeclareMathOperator{\Tr}{Tr}
\newcommand{\ket}[1]{\lvert #1\rangle}
\newcommand{\bra}[1]{\langle #1\rvert}
\newcommand{\id}{\mathbb I}
\newcommand{\C}{\mathbb C}
\newcommand{\Z}{\mathbb Z}
\newcommand{\HS}{\mathrm{HS}}
\newcommand{\Origin}{\mathcal O_Q}
\newcommand{\Ex}{\mathcal E}
\newcommand{\Cx}{\mathcal C}
\newcommand{\HH}{\mathsf H_2}
\newcommand{\Eop}{\mathsf E}
\newcommand{\Cop}{\mathsf C}
\begin{document}
\title{A spectral improvement of the Ashikhmin--Litsyn bound for quantum codes}
\author{Yu-Xuan Zhang}
\affiliation{School of Physics, Nankai University, Tianjin 300071, People's Republic of China}
\author{Jing-Ling Chen}
\email{chenjl@nankai.edu.cn}
\affiliation{Theoretical Physics Division, Chern Institute of Mathematics, Nankai University, Tianjin 300071, People's Republic of China}
\date{\today}

\begin{abstract}
The first asymptotic linear-programming bound of Ashikhmin and Litsyn limits the rate at which quantum information can be encoded at a prescribed relative distance. We prove a strictly smaller upper bound for general exact qubit codes throughout the positive-distance range of their result and obtain corresponding bounds in every integer local dimension. The multiplicity of an eigenspace on a Hamming annulus lowers the bound at zero error weight. To determine how much of this gain survives for quantum codes, we compare the dual function with its Fourier coefficients at every weight below the code distance. For qubits the comparison costs no exponential factor below a first contact at relative distance approximately $0.3153$. At the contact, increasing the shell radius suppresses the competing positive-weight contribution faster than it increases the shell entropy. The strict improvement therefore extends beyond the contact.
\end{abstract}
\maketitle

\section{Introduction}\label{sec:intro}
The distance of an error-correcting code determines how many errors it can detect. Increasing the distance restricts the number of messages that can be encoded at a given length~\cite{Hamming}. Delsarte expressed this restriction in terms of the distance distribution of a classical code and obtained upper bounds on its size by linear programming~\cite{Delsarte}. McEliece, Rodemich, Rumsey, and Welch (MRRW) used this method to bound the asymptotic rate at fixed relative distance~\cite{MRRW}. Friedman and Tillich recovered the first bound for binary linear codes from graph eigenvalues~\cite{FriedmanTillich}. Navon and Samorodnitsky obtained it for general binary codes by relating code size to the size and largest adjacency eigenvalue of a region in the Hamming cube~\cite{NavonSamorodnitsky}.

In quantum error correction, a message can be any state in a code subspace~\cite{Shor1995,Steane1996}. The corresponding question is how large this subspace can be if all errors of weight below a specified distance must be detectable. Shor and Laflamme introduced two quantum weight enumerators related by a MacWilliams identity~\cite{ShorLaflamme}. Rains developed their duality for general quantum codes~\cite{RainsWeight}. Ashikhmin and Litsyn used these relations to derive asymptotic upper bounds on the dimension of an exact quantum code~\cite{AshikhminLitsyn}. Their first bound for qubits has the same dependence on a Hamming shell radius as the first classical linear-programming bound, with a different normalization for the code dimension.

Both MRRW bounds for classical binary codes have recently been improved throughout the range of relative distances $0<\delta<1/2$. One proof assigns a spectral subspace to each codeword~\cite{OpenAICodes}. Alrabiah and Guruswami derived improved bounds using a binary-input channel with mixed quantum outputs~\cite{AlrabiahGuruswami}. Gay, Granha Jeronimo, and Liu refined the spectral construction by retaining transitions between additional subspaces~\cite{Honeycomb}.

In the spectral argument, a subspace of dimension $m$ on a support of size $D$ reduces the contribution to the logarithm of the code size bound from $\log_2 D$ to $\log_2(D/m)$~\cite{OpenAICodes,Barg,Samorodnitsky}. An exponentially large $m$ can therefore improve the rate bound. For a quantum code this gain enters at zero error weight. The quantum linear program takes a maximum over all errors of weight below the distance. Whether the gain survives this maximum is the question considered here.

We prove that the first Ashikhmin--Litsyn bound for general exact qubit codes can be strictly improved throughout its usual range $0<\delta\le0.3152$, where $\delta$ is the relative distance~\cite{AshikhminLitsyn,AshikhminLaiBrun}. The proof also gives explicit intervals of improvement in every local dimension $q\ge2$. We use an eigenspace on a thin Hamming annulus and compute the reduction in the bound due to its multiplicity. To pass from the origin to all detectable weights, we estimate the dual function and its Fourier coefficients using the same shell matrix elements. For qubits this comparison preserves the full multiplicity gain, on the exponential scale, up to $\delta\simeq0.3153$. There a positive weight first attains the same exponent as the origin in the unperturbed construction. Comparing the changes at the two weights shows how to continue the improvement.

This paper is organized as follows. Section~\ref{sec:dual} gives the quantum linear-programming bound and identifies the contribution of the spectral multiplicity. In Sec.~\ref{sec:construction} we construct the Hamming eigenspace and state the rate bound. Section~\ref{sec:comparison} proves it by comparing the error weights. Section~\ref{sec:contact} treats the first contact and the effect of varying the spectral parameters. We discuss the resulting rate estimates and the finite-density optimization in Sec.~\ref{sec:discussion}.

\section{The quantum linear-programming bound}\label{sec:dual}
Consider $n$ particles with $q$ internal levels each. A quantum code of dimension $K$ is a subspace of $(\C^q)^{\otimes n}$ with projector $P$ of rank $K$. The code detects an operator $E$ when $PEP$ is a scalar multiple of $P$. Its distance is at least $d$ if it detects every operator on fewer than $d$ particles~\cite{KnillLaflamme}. Write $K_q(n,d)$ for the largest such dimension, with value one if there is no nontrivial code. The rate at relative distance $\delta$ is
\begin{equation}
R_q(\delta)=\limsup_{n\to\infty}\frac{1}{n}\log_q K_q(n,\lceil\delta n\rceil).
\label{eq:rate}
\end{equation}
We use base-two logarithms for entropies and divide by $\log_2q$ to obtain rates in $q$-ary units.

\subsection{The comparison over detectable errors}
The tensor products of the single-particle Weyl operators form an orthogonal error basis for every integer $q$~\cite{Hostens}. Their labels $\vec{x}=(x_1,\ldots,x_n)$ belong to $\mathcal G=(\Z_q^2)^n$, and $\wt(\vec{x})$ is the number of nonzero coordinates. Put $Q=q^2$, $A=Q-1$, and $B=Q-2$. If $W_{\vec{x}}W_{\vec{y}}W_{\vec{x}}^\dagger=\chi_{\vec{x}}(\vec{y})W_{\vec{y}}$, we use the Fourier convention $f(\vec{x})=\sum_{\vec{y}}\widehat f(\vec{y})\overline{\chi_{\vec{x}}(\vec{y})}$.

For the projector $P$, let $\mathsf A_{\vec{x}}=|\Tr(PW_{\vec{x}})|^2/K^2$ and $\mathsf B_{\vec{x}}=\Tr(PW_{\vec{x}}PW_{\vec{x}}^\dagger)/K$. Expansion in the Weyl basis gives $\mathsf B_{\vec{x}}=(K/q^n)\sum_{\vec{y}}\chi_{\vec{x}}(\vec{y})\mathsf A_{\vec{y}}$. At a detectable error, $PW_{\vec{x}}P$ is scalar on the code and $\mathsf B_{\vec{x}}=\mathsf A_{\vec{x}}$. These two identities give the form of the quantum linear program we need~\cite{AshikhminLitsyn,LiXing}. Take a real function $f$ with $\widehat f\ge0$, $\widehat f(\vec{x})>0$ for $\wt(\vec{x})<d$, and $f(\vec{x})\le0$ for $\wt(\vec{x})\ge d$. Then
\begin{equation}
K\le q^{-n}\max_{\wt(\vec{x})<d}\frac{f(\vec{x})}{\widehat f(\vec{x})} .
\label{eq:LP}
\end{equation}
Indeed, writing $M$ for this maximum, the two enumerator identities give
\begin{align*}
Kq^n\sum_{\vec{y}}\widehat f(\vec{y})\mathsf A_{\vec{y}}
 &=\sum_{\vec{x}} f(\vec{x})\mathsf B_{\vec{x}}\\
 &\le\sum_{\wt(\vec{x})<d} f(\vec{x})\mathsf A_{\vec{x}}
 \le M\sum_{\vec{y}}\widehat f(\vec{y})\mathsf A_{\vec{y}} .
\end{align*}
The common sum is positive because $\mathsf A_{\vec{0}}=1$. This argument includes degenerate and nonadditive codes.

\subsection{The origin and the low-weight maximum}
Let $\nu$ be an even probability distribution on $\mathcal G$ whose Fourier sum $\Psi(\vec{x})=\sum_{\vec{y}}\nu(\vec{y})\overline{\chi_{\vec{x}}(\vec{y})}$ is nonnegative. Then $\Psi(\vec{0})=1$ and $\widehat\Psi=\nu$. Write $\mathcal A$ for Hamming adjacency, which sums a function over the $An$ neighbors of a label. The Fourier sum of the uniform distribution on these neighbors of zero is $t(\vec{x})=1-Q\wt(\vec{x})/(An)$. Thus $f=(t-s)\Psi$, with $s=1-Qd/(An)$, has the required sign above the distance. Its Fourier coefficients are
\begin{equation}
\widehat f=\frac{\mathcal A\nu}{An}-s\nu .
\label{eq:dual}
\end{equation}
If $\mathcal A\nu/(An)\ge\Lambda\nu$ with $\Lambda>s$, these coefficients are positive wherever $\nu>0$.

Suppose $\nu$ is positive below the distance. Substituting Eq.~\eqref{eq:dual} into Eq.~\eqref{eq:LP} gives
\begin{equation}
K\le\frac{1}{q^n}\frac{1-s}{\Lambda-s}
       \frac{1}{\nu(\vec{0})}
       \max_{\wt(\vec{x})<d}\frac{\nu(\vec{0})\Psi(\vec{x})}{\nu(\vec{x})} .
\label{eq:transfer}
\end{equation}
The last factor compares two functions normalized to one at zero. In fact, $\Psi/[Q^n\nu(\vec{0})]$ is a probability distribution, and its Fourier sum is $\nu/\nu(\vec{0})$ by inversion. When the maximum is $2^{o(n)}$ and the spectral slack $\Lambda-s$ stays bounded away from zero, $-\log_2\nu(\vec{0})$ alone determines the exponential contribution from these functions. The same bound holds for any even probability kernel $\eta$ with $\eta*\nu\ge\Lambda\nu$ if its Fourier sum $t$ is at most $s$ outside the detectable set. Equation~\eqref{eq:dual} then reads $\widehat f=\eta*\nu-s\nu$.

To construct $\nu$, take positive matrices $F(\vec{u})$ of size $m$ with $\sum_{\vec{u}}F(\vec{u})=\id_m$ and set
\begin{equation}
\begin{aligned}
\nu(\vec{y})&=\frac1m\sum_{\vec{u}}\Tr[F(\vec{u})F(\vec{u}+\vec{y})],\\
\Psi(\vec{x})&=\frac1m\left\|\sum_{\vec{u}}\chi_{\vec{x}}(\vec{u})F(\vec{u})\right\|_{\HS}^2 .
\end{aligned}
\label{eq:frame}
\end{equation}
Expansion of the square gives the Fourier relation above. To interpret $\nu$, give label $\vec{u}$ probability $\mu(\vec{u})=\Tr F(\vec{u})/m$ and encode it in the state $F(\vec{u})/\Tr F(\vec{u})$ whenever $\mu(\vec{u})>0$. The average state is $\id_m/m$, so the pretty good measurement has effects $F(\vec{v})$~\cite{HausladenWootters}. The joint probability of preparing $\vec{u}$ and observing $\vec{v}$ is then $\Tr[F(\vec{u})F(\vec{v})]/m$. Equation~\eqref{eq:frame} gives the distribution of $\vec{v}-\vec{u}$, with return probability $\nu(\vec{0})$ for a correct guess~\cite{BarnumKnill,BertaColesWehner}. Equivalently, $-\log_2\nu(\vec{0})=\HH^\downarrow(U|\mathsf B)$, the conditional collision entropy of the label given the encoded state, evaluated with its actual marginal. For rank-one effects $\Tr[F(\vec{u})^2]=(\Tr F(\vec{u}))^2$, so $\nu(\vec{0})=m\sum_{\vec{u}}\mu(\vec{u})^2$. Equal weights on $D$ labels give $\nu(\vec{0})=m/D$.

To obtain the spectral inequality, restrict $\mathcal A$ to a connected support and let $\vec{\phi}=(\phi(\vec{u}))_{\vec{u}}$ be its positive Perron vector of eigenvalue $\lambda_0$. Choose $m$ independent eigenfunctions of eigenvalue $\lambda$, collect their values at $\vec{u}$ in $\vec{v}(\vec{u})$, and normalize the basis so that $F(\vec{u})=\vec{v}(\vec{u})\vec{v}(\vec{u})^\dagger/\phi(\vec{u})$ sums to $\id_m$. Extend $F$ by zero off the support. On the support the weights $\mathcal A_{\vec{u},\vec{w}}\phi(\vec{w})/[\lambda_0\phi(\vec{u})]$ define a transition matrix $P_{\vec{\phi}}$. For $\vec{z}(\vec{u})=\vec{v}(\vec{u})/\phi(\vec{u})$, the eigenvalue equation gives $P_{\vec{\phi}}\vec{z}=(\lambda/\lambda_0)\vec{z}$. Hence
\begin{multline*}
(\mathcal AF)(\vec{u})-\frac{\lambda^2}{\lambda_0}F(\vec{u})\\
=\lambda_0\phi(\vec{u})\big[P_{\vec{\phi}}(\vec{z}\vec{z}^\dagger)
 -(P_{\vec{\phi}}\vec{z})(P_{\vec{\phi}}\vec{z})^\dagger\big](\vec{u})\succeq0 .
\end{multline*}
The last matrix is a positive multiple of a conditional covariance. Outside the support $F=0$ and $\mathcal AF\succeq0$, so the inequality holds everywhere. Multiplying by $F(\vec{u}+\vec{y})$, taking the trace, and summing over $\vec{u}$ gives $\mathcal A\nu/(An)\ge\Lambda\nu$ with $\Lambda=\lambda^2/(An\lambda_0)$. We can now choose a spectral subspace and compute both its contribution at the origin and the remaining maximum in Eq.~\eqref{eq:transfer}.

\section{The Hamming construction and the rate bound}\label{sec:construction}
A word of weight $i$ in the Weyl label space has $i$ nonzero coordinates, with $A$ choices at each one. The shell of weight $i$ therefore contains $D_i=\binom ni A^i$ words. If $i/n\to a$, its entropy per coordinate is $H_Q(a)=h_2(a)+a\log_2 A$, where $h_2$ is the binary entropy. In the first linear-programming bound the radius corresponding to distance $\delta$ is $a_0=g_Q(\delta)$, with
\begin{equation}
\begin{aligned}
g_Q(a)&=\frac{A-Ba-2\sqrt{Aa(1-a)}}{Q},\\
L_q(\delta)&=\frac{H_Q(g_Q(\delta))}{\log_2q}-1 .
\end{aligned}
\label{eq:benchmark}
\end{equation}
The function $g_Q$ is a decreasing involution on $[0,A/Q]$. For qubits, $L_2$ is the first Ashikhmin--Litsyn expression. We use $L_q$ as the comparison function in general local dimension.

To introduce multiplicity while keeping the radius well defined, take the shells with $L-\lfloor n^{2/3}\rfloor\le i\le L$, where $L/n\to a$. They all have entropy $nH_Q(a)+o(n)$. On one coordinate write $\vec{e}_0$ for the basis vector of the zero symbol, $\vec{s}_1=A^{-1/2}\sum_{u\ne0}\vec{e}_u$, and $U$ for the space of vectors supported on the nonzero symbols and orthogonal to $\vec{s}_1$. Its dimension is $B$. Choose $k$ coordinates on which to use $U$, and take the uniform radial vector on the remaining coordinates at the weight needed to reach shell $i$. We call the chosen coordinates anchors. Their positions and their $B$ internal directions give
\begin{equation}
m_k=\binom nk B^k .
\label{eq:sector}
\end{equation}
Different choices of anchor positions give orthogonal spaces. In each copy the adjacency between shells is the same tridiagonal matrix. The top eigenvector of that matrix therefore gives an eigenspace of dimension $m_k$ on the annulus.

Put $k/n\to b$. The Perron eigenvalue of the annulus and the eigenvalue in this subspace, divided by $n$, tend respectively to
\begin{equation}
\begin{aligned}
\rho_0(a)&=Ba+2\sqrt{Aa(1-a)},\\
\rho_b(a)&=Ba-Ab+2\sqrt{A(a-b)(1-a)} .
\end{aligned}
\label{eq:rho}
\end{equation}
For the field in Eq.~\eqref{eq:frame}, the squared shell weights change the return probability only by a subexponential factor. Thus the limiting spectral ratio and the origin entropy per coordinate are
\begin{equation}
\begin{aligned}
\Lambda_Q(a,b)&=\frac{\rho_b(a)^2}{A\rho_0(a)},\\
\Origin(a,b)&=\lim_{n\to\infty}-\frac1n\log_2\nu(\vec{0})\\
 &=H_Q(a)-h_2(b)-b\log_2 B .
\end{aligned}
\label{eq:profiles}
\end{equation}
Appendix~\ref{app:shells} derives both expressions from the shell eigenvectors. At $b=0$ the spectral equality $\Lambda_Q(a,0)=1-Q\delta/A$ gives $a=a_0$. For small $b>0$, write $a_q(\delta,b)$ for the nearby solution of
\begin{equation}
\Lambda_Q(a_q(\delta,b),b)=1-Q\delta/A .
\label{eq:implicit}
\end{equation}
The radius increases by $O(b)$. The increase in shell entropy is of this same order, whereas the multiplicity subtracts $h_2(b)+b\log_2B$. Whether this reduction in Eq.~\eqref{eq:profiles} improves the rate depends on the low-weight maximum.

\begin{theorem}\label{thm:lossless}
For every integer $q\ge2$ and compact interval $I\subset(0,\bar d_q)$, there is a $b_I>0$ such that for $\delta\in I$ and $0<b<b_I$,
\begin{equation}
\begin{aligned}
R_q(\delta)&\le\frac{\Origin(a_q(\delta,b),b)}{\log_2q}-1\\
 &=L_q(\delta)-\frac{h_2(b)}{\log_2q}+O_I(b).
\end{aligned}
\label{eq:lossless}
\end{equation}
For $q=2,3,4$, $\bar d_q$ is the first positive-weight contact of the $k=0$ construction, located in Appendix~\ref{app:geometry} and given in Table~\ref{tab:contacts}. For $q\ge5$ one may take $\bar d_q=g_Q(\vartheta_q)$, where
\begin{equation}
\vartheta_q=\frac{(q-1)(q+2+\sqrt{q+1})}
                  {2q(q+2+2\sqrt{q+1})}.
\label{eq:threshold}
\end{equation}
\end{theorem}

The density $b$ is fixed before the length tends to infinity. Since $h_2(b)/b\to\infty$ as $b\downarrow0$, Eq.~\eqref{eq:lossless} gives $R_q(\delta)<L_q(\delta)$ at every distance in the stated open interval. In particular, the strict comparison for qubits includes the usual endpoint $0.3152$ of the Ashikhmin--Litsyn bound. The radius adjustment has already been included in $a_q(\delta,b)$. The rest of the proof shows that taking the maximum in Eq.~\eqref{eq:transfer} incurs no further exponential cost.

\section{Comparison at low error weights}\label{sec:comparison}
Let $j=\wt(\vec{x})$, $\xi=j/n$, and denote the two exponents to be compared by
\begin{equation}
e_n(\xi)=\frac1n\log_2\Psi(\vec{x}),\qquad
d_n(\xi)=\frac1n\log_2\frac{\nu(\vec{x})}{\nu(\vec{0})} .
\label{eq:exponents}
\end{equation}
For this construction both quantities depend only on $j$ and vanish at $j=0$. At a prescribed relative weight $\xi$ we take $j/n\to\xi$ in the length limit. To bound the maximum of $e_n-d_n$, first take $k=0$. On one shell the numerator $\Psi(\vec{x})$ is the squared average of the character $\chi_{\vec{x}}$ over that shell. The normalized displacement $\nu(\vec{x})/\nu(\vec{0})$ is the fraction of the shell that remains in the shell after translation by $\vec{x}$. Denote their respective limiting exponents by $E_Q(a,\xi)$ and $D_Q(a,\xi)$ in the region before the character's turning point $\xi=g_Q(a)$, and write
\begin{equation}
\Gamma_Q(a,\xi)=E_Q(a,\xi)-D_Q(a,\xi).
\label{eq:gap}
\end{equation}
We use the continuous value of $E_Q$ at the turning point. The character average and the intersection count give the two saddle equations in Appendix~\ref{app:stability}. At $a_0=g_Q(\delta)$ they imply two estimates. The gap has strictly negative slope at zero, and its maximum on $[\xi_0,\delta]$ is strictly negative for every $0<\xi_0\le\delta$. These signs hold uniformly on compact distance intervals in Theorem~\ref{thm:lossless}. Appendix~\ref{app:geometry} derives them and locates the first distance where equality occurs at a positive weight.

Now consider $k>0$. The character and the displacement must be evaluated for the same annular field. A matrix element between shells of weights $i$ and $\ell$ occurs on both sides of the comparison. Its squared modulus contributes to the displacement, and the Fourier sum of those contributions gives the character. Computing the two quantities from the same kernel retains the terms with $i\ne\ell$ (Appendix~\ref{app:shells}). On a single shell the character sum groups naturally according to how many anchors change position. On the displacement side the sum is a weighted sum of squared amplitudes. An individual amplitude can vanish, which is why the lower estimate uses the whole sum.

Near zero weight the estimates obtained this way are, uniformly as $n\to\infty$,
\begin{equation}
\begin{aligned}
e_n(\xi)&\le E_Q(a,\xi)+C\xi\sqrt b+o_n(1),\\
d_n(\xi)&\ge D_Q(a,\xi)-Cb\xi-o_n(1).
\end{aligned}
\label{eq:local-bounds}
\end{equation}
The factor $\xi$ matters because the rank-one gap itself vanishes at zero. At weights bounded away from zero, an error tending to zero with $b$ suffices. On a compact interval before the turning point where the maximizing pair of shell words satisfies the positivity condition in Appendix~\ref{app:stability}, the more precise result is
\begin{equation}
\begin{aligned}
e_n(\xi)&=E_Q(a,\xi)+h_2(b)+O(b)+o_n(1),\\
d_n(\xi)&=D_Q(a,\xi)+O(b)+o_n(1).
\end{aligned}
\label{eq:interior-bounds}
\end{equation}
The term $h_2(b)$ will be needed when the rank-one gap closes. Appendix~\ref{app:stability} gives the uniform estimates, including at the turning point.

\subsection{Proof of Theorem~\ref{thm:lossless}}
At $b=0$, $\partial_a\Lambda_Q(a,0)>0$ for $0<a<A/Q$ and $\partial_b\Lambda_Q(a,0)<0$. Equation~\eqref{eq:implicit} therefore gives $a_q(\delta,b)=a_0+O_I(b)>a_0$. Use the slightly larger radius $a=a_q(\delta,b)+\epsilon$, $0<\epsilon\le b^2$, so the spectral inequality is strict at fixed $b,\epsilon$.

Near zero, the rank-one gap is at most $-c\xi$ uniformly on $I$, and the small increase in radius preserves this estimate. Subtraction in Eq.~\eqref{eq:local-bounds} gives
\[
e_n(\xi)-d_n(\xi)\le(-c+C\sqrt b+Cb)\xi+o_n(1).
\]
This also covers errors whose weight grows sublinearly in $n$. In the rest of the interval the fixed negative gap absorbs the perturbation for all sufficiently small $b$. The continuous upper estimate for the character in Appendix~\ref{app:stability} covers the short interval crossed by its turning point as $a$ increases. Thus $\max_{j<d}(e_n-d_n)=o_n(1)$, since the value at $j=0$ is zero. Taking the length limit in Eq.~\eqref{eq:transfer} and then letting $\epsilon$ decrease to zero gives Eq.~\eqref{eq:lossless}.

\subsection{Exchange of anchor positions}
The estimates near zero and at fixed positive weight come from the same counting problem. Suppose $r$ of the $j$ affected positions lie in the original anchor set. To change $t$ anchors one chooses $t$ of these $r$ positions and $t$ of the other $j-r$ positions. This gives the factor $\binom rt\binom{j-r}t$ in the shell character. At positive weight the term which exchanges all $k$ anchors already contributes $h_2(b)+O(b)$ to the exponent. As $\xi$ decreases, fewer such choices are available.

The transition occurs at $\xi=u\sqrt b$. Here the leading exchange sum reduces to selecting equally many positions from sets of sizes $k$ and $j$, with a weight for each exchange. For general sizes $k\le M$ the corresponding sum is
\begin{equation}
Z_{k,M}(x)=\sum_{t=0}^k\binom kt\binom Mt
                    \left(\frac{kx^2}{M}\right)^t .
\label{eq:reservoir}
\end{equation}
If $k\to\infty$ and $k/M\to0$, the fraction $v=t/k$ that contributes at exponential order satisfies $v^2=x^2(1-v)$. In the Hamming construction $M=j$ and $x=c_a u$, with
\[
c_a=\frac{a\sqrt{\beta_Q}}{(1-Qa/A)^2},\qquad
\beta_Q=\frac{Q^2 B}{A^4}.
\]
Writing $\mathfrak F(x)=[v-\ln(1-v)]/\ln2$ for the limiting value of $k^{-1}\log_2Z_{k,M}(x)$, we obtain
\begin{equation}
\lim_{n\to\infty}e_n(u\sqrt b)
 =E_Q(a,u\sqrt b)+b\,\mathfrak F(c_a u)+o(b).
\label{eq:crossover}
\end{equation}
The displacement exponent differs from $D_Q(a,u\sqrt b)$ by $o(b)$ in the same successive limits. For small $x$, $v=x+O(x^2)$, and the change in Eq.~\eqref{eq:crossover} is proportional at leading order to $\xi\sqrt b$. As $x$ grows, $v$ approaches one. The same finite set of anchors then accounts for the saturation of the exchange. Appendix~\ref{app:exchange} derives Eq.~\eqref{eq:crossover} and the fixed-weight return formula. The fixed positive-weight estimate in Eq.~\eqref{eq:interior-bounds} follows directly from the shell sum.

\section{Continuation at a positive error weight}\label{sec:contact}
At the first contact $\delta_c$ the rank-one gap is zero both at $\xi=0$ and at an interior weight $\xi_c>0$. Equation~\eqref{eq:interior-bounds} shows what changes in the argument. A small anchor density lowers the origin entropy by $h_2(b)+O(b)$, and raises the normalized logarithmic ratio at $\xi_c$ by $h_2(b)+O(b)$. Along the radius branch of Eq.~\eqref{eq:implicit}, the radius changes by only $O(b)$. To keep a gain of order $h_2(b)$ we must also change the radius on that scale.

\subsection{The change of radius}
Write $a_c=g_Q(\delta_c)$. Suppose $\Gamma_Q(a_c,\xi)\le0$ on $[0,\delta_c]$, with negative slope at zero, a strict gap at $\delta_c<\min(1,2a_c)$, and finitely many positive zeros $\xi_\ell$ in the interior. Assume that Eq.~\eqref{eq:interior-bounds} holds near each zero. There are two rates of change to compare,
\begin{equation}
H_a=H'_Q(a_c),\qquad
G=\min_\ell\{-\partial_a\Gamma_Q(a_c,\xi_\ell)\}.
\label{eq:slopes}
\end{equation}
Increasing the radius raises the shell entropy at rate $H_a$ and lowers the gap at every contact at rate at least $G$. The contacts for $q=2,3,4$ all satisfy $G>H_a>0$ (Table~\ref{tab:contacts}).

Put $h_b=h_2(b)$ and take $a_b=a_c+\tau h_b$, $\delta_b=\delta_c+\sigma h_b$, with fixed $\tau>0$ and $\sigma\ge0$. At a contact with $G_\ell=-\partial_a\Gamma_Q(a_c,\xi_\ell)$, Eq.~\eqref{eq:interior-bounds} gives
\begin{align*}
e_n(\xi_\ell)-d_n(\xi_\ell)
 &=\Gamma_Q(a_b,\xi_\ell)+h_b+O(b)+o_n(1)\\
 &=(1-G_\ell\tau)h_b+o(h_b)+o_n(1).
\end{align*}
The first-order value is also the maximum to this accuracy in a small neighborhood of $\xi_\ell$. Indeed, as $b\downarrow0$ every maximizing sequence there approaches $\xi_\ell$, and the expansion in the radius is uniform. The origin contributes zero. The gap elsewhere stays strict away from zero, including on the short interval added at $\delta_c$.

For the dual $f_n=(t-s_n)\Psi$, $s_n=1-Q\lceil n\delta_b\rceil/(An)$, define the excess over its origin value by
\[
X_n=\frac1n\log_2
 \frac{\displaystyle\max_{\wt(\vec{x})<\lceil n\delta_b\rceil} f_n(\vec{x})/\widehat f_n(\vec{x})}
      {f_n(\vec{0})/\widehat f_n(\vec{0})} .
\]
Equation~\eqref{eq:dual} bounds its upper length limit by $[1-G\tau]_+h_b+o(h_b)$, where $[y]_+=\max(y,0)$. For the reverse estimate, take $j/n\to\xi_\ell$ at a contact attaining $G$. The adjacency in $\widehat f_n$ uses only weights $j-1,j,j+1$, whose normalized displacement exponents are at most $D_Q(a_b,\xi_\ell)+o_n(1)$. The factor $t-s_n$ stays positive at this interior weight and $\widehat f_n(\vec{0})$ is bounded above and below by positive multiples of $\nu(\vec{0})$ for fixed $b$. Thus both $\liminf_n X_n$ and $\limsup_n X_n$ have the same first-order value.

The spectral slack is positive at $a_b$, since $\partial_a\Lambda_Q(a_c,0)>0$ and $b=o(h_b)$. Adding the origin contribution to the excess gives
\begin{equation}
\begin{aligned}
R_q(\delta_b)\le {}& L_q(\delta_c)\\
 &+\frac{H_a\tau-1+[1-G\tau]_+}{\log_2q}\,h_b+o(h_b).
\end{aligned}
\label{eq:contact-rate}
\end{equation}
For $\tau>1/G$ the contact neighborhoods have strictly negative excess, so $\limsup_n X_n=0$ for small fixed $b$. The best first-order rate in Eq.~\eqref{eq:contact-rate} occurs at $\tau=1/G$, leaving the fraction $\kappa=1-H_a/G$ of the gain. Finally $g_Q$ is an involution, so $L'_q(\delta_c)=-H_a/(|g'_Q(a_c)|\log_2q)$. Comparing at the new distance gives
\begin{equation}
R_q(\delta_b)\le L_q(\delta_b)
 -\frac{\kappa-H_a\sigma/|g'_Q(a_c)|}{\log_2 q}\,h_b+o(h_b).
\label{eq:continuation}
\end{equation}
The improvement extends past $\delta_c$ for every $0<\sigma<\kappa |g'_Q(a_c)|/H_a$. For qubits the strict estimates in Appendix~\ref{app:geometry} permit the explicit radius shift $\tau=181/1000$. For all sufficiently small fixed $b>0$,
\begin{equation}
R_2(\delta_{c,2}+h_b/10)
 \le L_2(\delta_{c,2}+h_b/10)-h_b/50 .
\label{eq:qubit}
\end{equation}

\begin{table}[tb]
\caption{First interior contact for each of the three smallest local dimensions. Here $a_c$ is the shell radius, $\delta_c$ the relative code distance, $\xi_c$ the relative error weight at contact, and $\kappa$ the fraction of the leading gain retained at $\delta_c$. Values are rounded.}
\label{tab:contacts}
\begin{ruledtabular}
\begin{tabular}{ccccc}
$q$ & $a_c$ & $\delta_c$ & $\xi_c$ & $\kappa$\\
\hline
2 & 0.1900 & 0.3153 & 0.3009 & 0.3344\\
3 & 0.2562 & 0.4152 & 0.3973 & 0.3333\\
4 & 0.2910 & 0.4630 & 0.4438 & 0.3319
\end{tabular}
\end{ruledtabular}
\end{table}

\subsection{Several contacts and several parameters}
The comparison in Eq.~\eqref{eq:slopes} also applies when the spectral construction has more than one adjustable parameter. To see what changes, let $\vec{\theta}$ denote these parameters and let $\vec{v}$ be an admissible direction in which to change them by $\epsilon\vec{v}$. Write $\vec{h}$ for the gradient of the origin entropy and $\vec{g}_\xi=-\nabla_{\vec{\theta}}\Gamma(\vec{\theta},\xi)$ for the rate of suppression at a contact $\xi$. Suppose that adding multiplicity saves $\epsilon$ at the origin and adds $\epsilon$ to the gap at each positive contact to leading order. As above, the origin supplies a zero excess. On a compact set $\mathcal Z$ of positive contacts, the largest excess comes from a contact with the smallest $\vec{g}_\xi\cdot\vec{v}$. The first-order change in the sum of origin entropy and excess is therefore
\begin{equation}
J(\vec{v})=\vec{h}\cdot\vec{v}-1+
       \big[1-\inf_{\xi\in\mathcal Z}\vec{g}_\xi\cdot\vec{v}\big]_+ .
\label{eq:cost}
\end{equation}
A direction yields a strict improvement after rescaling precisely when its suppression rate at every contact exceeds its cost at the origin. When varying only the Hamming radius this criterion gives $G>H_a$. With several parameters the same direction must work at all competing weights. Appendix~\ref{app:directions} gives the uniform hypotheses, the result for unequal contact penalties, and the corresponding minimization over admissible directions.

\section{Discussion}\label{sec:discussion}
We have obtained an asymptotic upper bound on the rate of general exact quantum codes. For qubits it strictly improves the first Ashikhmin--Litsyn bound throughout its standard positive-distance range. For arbitrary integer local dimension the interval in Theorem~\ref{thm:lossless} follows from the same Hamming construction. At the first interior contacts in Table~\ref{tab:contacts}, Eq.~\eqref{eq:continuation} gives a further interval of strict improvement.

The proof compares two effects of using an eigenspace of large dimension. Its multiplicity increases the return probability of the spectral measurement, lowering the origin contribution to the rate bound. Errors of positive weight can exchange anchor positions, which increases their contribution to the maximum in the quantum linear program. A strict gap in the rank-one comparison absorbs this increase. At a contact the two contributions have to be estimated at the same order. Their changes with the shell radius then determine how much of the multiplicity can still be used.

At $\delta=1/4$ the comparison can be made directly. Here $a_0=1/4$, $L_2(1/4)=1-\tfrac12\log_2 3$, and Eq.~\eqref{eq:implicit} gives $a_2(1/4,b)=1/4+3b+O(b^2)$. Expanding Eq.~\eqref{eq:lossless} yields
\begin{equation}
\begin{aligned}
R_2(1/4)\le {}&1-\tfrac12\log_2 3-b\log_2(1/b)\\
 &+(6\log_2 3-1-\log_2 e)b+O(b^2).
\end{aligned}
\label{eq:example}
\end{equation}
The first linear-programming value at this distance is already below the quantum Hamming value $1-H_4(1/8)$ and the Singleton value $1/2$~\cite{AshikhminLaiBrun,ZhangChenBinary,ZhangChenQary}. Equation~\eqref{eq:example} thus lowers the minimum of these three bounds. Independent bounds also restrict the relevant distance range. In particular, the asymptotic relative distance of nontrivial codes is at most $(1-q^{-2})/2$ in general and $(3-\sqrt3)/4$ for qubits~\cite{Rains2003}.

For small fixed anchor density the improvement is strict, while its optimal size depends on the choice of radius and density. Appendix~\ref{app:finite} gives an explicit variational bound at finite density. Its denominator involves the total squared coefficient over physical shell pairs. Additional information about these coefficients can strengthen the estimate, including when some of the individual amplitudes vanish. Optimizing that bound would give quantitative rate curves beyond the small-density expansion. Another possibility is to vary several spectral parameters and use Eq.~\eqref{eq:cost} to determine which changes reduce the contributions at all active error weights.

\begin{acknowledgments}
This work is supported by the Quantum Science and Technology--National Science and Technology Major Project (Grant No.~2024ZD0301000), and the National Natural Science Foundation of China (Grant No.~12275136).
\end{acknowledgments}

\onecolumngrid
\appendix

\section{Matrix elements on Hamming shells}\label{app:shells}
\subsection{The annular eigenspace}
On one coordinate the Hamming adjacency sends $\vec{e}_0$ to $\sqrt A\,\vec{s}_1$, sends $\vec{s}_1$ to $\sqrt A\,\vec{e}_0+B \vec{s}_1$, and acts as $-\id$ on $U$. Let $E_k=\bigoplus_{|S|=k}U^{\otimes S}$. For each $i\ge k$, the isometry $\Phi_{i,k}$ into shell $i$ places a vector of $U^{\otimes S}$ on the coordinates in $S$ and the normalized uniform radial vector of weight $i-k$ on $S^c$. Two different sets $S$ are orthogonal because a color direction on one of their differing coordinates is orthogonal to both $\vec{e}_0$ and $\vec{s}_1$. The dimensions of the source and target are $m_k$ in Eq.~\eqref{eq:sector} and $D_i=\binom ni A^i$, respectively.

For a fixed vector in $E_k$, the diagonal adjacency contribution on shell $i$ is $B(i-k)-k$. Changing one radial coordinate from zero to nonzero couples adjacent shells. Normalization of the two uniform vectors gives
\begin{equation}
J^{(k)}_{ii}=Bi-Ak,\qquad
J^{(k)}_{i,i+1}=\sqrt{A(i-k+1)(n-i)} .
\label{eq:jacobi}
\end{equation}
Restrict this tridiagonal matrix to $I_n=\{L-\lfloor n^{2/3}\rfloor,\ldots,L\}$, with $L/n\to a$, $k/n\to b<a$. The entries of $J^{(k)}/n$ converge uniformly there to $Ba-Ab$ on the diagonal and $\sqrt{A(a-b)(1-a)}$ on each adjacent diagonal. The maximum row sum bounds the top eigenvalue from above. For a lower bound use the positive sine vector on an interval of $\ell$ shells, where $\ell\to\infty$ and $\ell\le |I_n|$. Its normalized nearest-neighbor correlation is $\cos[\pi/(\ell+1)]$. The two bounds give $\lambda_k/n\to\rho_b(a)$ in Eq.~\eqref{eq:rho}. The annulus is connected and permutations of coordinates and nonzero symbols act transitively on each shell. Its positive Perron vector is therefore radial, giving the $k=0$ value $\lambda_0/n\to\rho_0(a)$.

The entry of $\vec{\phi}$ at a word $\vec{u}$ on shell $i$ is $\phi(\vec{u})=h_i/\sqrt{D_i}$. If $\vec{g}=(g_i)_{i\in I_n}$ is the positive normalized top eigenvector of $J^{(k)}$, then $W=\sum_{i\in I_n}g_i\Phi_{i,k}$ embeds $E_k$ isometrically into the $\lambda_k$ eigenspace. Put
\[
\sigma_0=\sum_{i\in I_n}\frac{g_i^2\sqrt{D_i}}{h_i},\qquad
p_i=\frac{g_i^2\sqrt{D_i}}{\sigma_0 h_i}.
\]
The matrices $F(\vec{u})=W^\dagger\ket{\vec{u}}\bra{\vec{u}} W/[\sigma_0\phi(\vec{u})]$ sum to $\id_{m_k}$. For $D_{\vec{x}}(\vec{u},\vec{u})=\chi_{\vec{x}}(\vec{u})$ let $T_{i,k}(\vec{x})=\Phi_{i,k}^\dagger D_{\vec{x}}\Phi_{i,k}$. Transitivity makes the diagonal of $\Phi_{i,k}\Phi_{i,k}^\dagger$ constant on the shell, with value $m_k/D_i$. Equation~\eqref{eq:frame} now gives
\begin{equation}
\Psi(\vec{x})=\frac1{m_k}\left\|\sum_{i\in I_n}p_iT_{i,k}(\vec{x})\right\|_{\HS}^2,\qquad
\nu(\vec{0})=m_k\sum_{i\in I_n}\frac{p_i^2}{D_i}.
\label{eq:mixture}
\end{equation}
All $D_i$ have the same exponential order and $\sum_i p_i=1$, so $|I_n|^{-1}\le\sum_i p_i^2\le1$. Taking logarithms yields the second identity in Eq.~\eqref{eq:profiles}. For strict spectral slack, the origin exponent of the actual ratio $f(\vec{0})/\widehat f(\vec{0})$ is the same. Indeed $f(\vec{0})=1-s$, and Cauchy--Schwarz in Eq.~\eqref{eq:frame} gives $\nu(\vec{y})\le\nu(\vec{0})$, whence $(\Lambda-s)\nu(\vec{0})\le\widehat f(\vec{0})\le(1-s)\nu(\vec{0})$.

\subsection{One kernel for the displacement and its Fourier sum}
Let $P_{i,\ell;k}=\Phi_{i,k}\Phi_{\ell,k}^\dagger$ and $C_i=\binom{n-k}{i-k}$. Before taking a Fourier transform, the contribution of a pair of shells to $\nu(\vec{x})$ is proportional to $\sum_{|\vec{u}|=i,\ |\vec{u}+\vec{x}|=\ell}|P_{i,\ell;k}(\vec{u},\vec{u}+\vec{x})|^2$. Here $|\vec{u}|$ denotes Hamming weight. In one copy of the construction use the polynomial matrix
\[
\mathsf L(x,y;t)=(\vec{e}_0+x\vec{s}_1)(\vec{e}_0+y\vec{s}_1)^\dagger+xy\,tP_U ,
\]
where $P_U$ projects onto $U$ and the formal variables are real. The coefficient of $x^iy^\ell t^k$ in $\mathsf L^{\otimes n}$ is $\sqrt{C_i C_\ell}\,P_{i,\ell;k}$. To square a matrix element, use two copies with anchor variables $u,v$ and keep the shell degrees in $X,Y$. A zero displacement has either two zero symbols or two equal nonzero symbols. A nonzero displacement has one zero symbol or two unequal nonzero symbols. The entries of $P_U$ on equal and unequal nonzero symbols are $B/A$ and $-1/A$. Thus the two local factors and their Fourier transforms are
\begin{equation}
\begin{aligned}
\Cop_0&=1+\frac{XY}{A}(1+Bu)(1+Bv),&
\Cop_1&=\frac{X+Y}{A}+\frac{BXY}{A^2}(1-u)(1-v),\\
\Eop_0&=\Cop_0+A\Cop_1,&
\Eop_1&=\Cop_0-\Cop_1 .
\end{aligned}
\label{eq:kernel}
\end{equation}
A nontrivial character sums to $-1$ on the nonzero symbols, giving the second row. For $j=\wt(\vec{x})$ the full shell contributions are
\begin{align}
\sum_{|\vec{u}_0|=i,\ |\vec{u}_0+\vec{x}|=\ell}|P_{i,\ell;k}(\vec{u}_0,\vec{u}_0+\vec{x})|^2
 &=\frac{[X^iY^\ell u^kv^k]\,\Cop_0^{n-j}\Cop_1^j}{C_i C_\ell},\label{eq:master-C}\\
\Tr[T_{i,k}(\vec{x})T_{\ell,k}(\vec{x})^\dagger]
 &=\frac{[X^iY^\ell u^kv^k]\,\Eop_0^{n-j}\Eop_1^j}{C_i C_\ell}.\label{eq:master-E}
\end{align}
Multiplication by $p_i p_\ell/m_k$ and summation over $i,\ell$ give $\nu(\vec{x})$ and $\Psi(\vec{x})$, respectively.

For the estimates it is useful to expand the same-shell terms. The radial character on shell $i$ and the two local color weights are
\begin{equation}
\varphi_i^{(Q,n)}(j)=
\frac{[z^i](1+Az)^{n-j}(1-z)^j}{A^i\binom ni},\qquad
\alpha_Q=1-\frac{Q}{A^2},\qquad \beta_Q=\frac{Q^2B}{A^4}.
\label{eq:spherical}
\end{equation}
Fix an input anchor set $S$. Let $r$ be the number of its positions affected by $\vec{x}$, and $t$ the number of those anchors replaced by positions in the affected complement. Choosing the intersection and the $t$ positions on each side gives the first three binomial factors below. The retained anchors and the exchanged pairs have squared weights $\alpha_Q$ and $\beta_Q$. These follow by restricting the character modulation to $U$ and to its complementary direction $\vec{s}_1$, using $\langle \vec{s}_1,D_{x_p}\vec{s}_1\rangle=-1/A$ on a coordinate $p$ with $x_p\ne0$. The radial vector must occupy the $t$ gained positions. Conditioning on them leaves the radial character of the remaining positions. Summing the squared norms of the orthogonal output blocks gives
\begin{equation}
\begin{aligned}
\Ex_{i,k}(j)&:=\frac{\|T_{i,k}(\vec{x})\|_{\HS}^2}{m_k}\\
 &=\sum_r\frac{\binom jr\binom{n-j}{k-r}}{\binom nk}
   \sum_t\binom rt\binom{j-r}t\,\alpha_Q^{r-t}\beta_Q^t
   \left[\frac{\binom{n-k-t}{i-k-t}}{\binom{n-k}{i-k}}\,
      \varphi_{i-k-t}^{(Q,n-k-t)}(j-r-t)\right]^2 .
\end{aligned}
\label{eq:migration}
\end{equation}
Here $\max(0,k+j-n)\le r\le\min(k,j)$ and $0\le t\le\min(r,j-r,i-k)$. The radial character is real also for complex Weyl characters, by the sum over the nonzero symbols in Eq.~\eqref{eq:spherical}.

For the displacement, let $c,h$ count positions at which both symbols are nonzero and are equal or unequal, respectively. Directly from the entries of $P_U$,
\begin{equation}
P_{i,\ell;k}(\vec{u},\vec{v})=\frac{\mathcal A_k(c,h)}{A^{(i+\ell)/2}\sqrt{C_iC_\ell}},\qquad
\mathcal A_k(c,h)=[z^k](1+Bz)^c(1-z)^h .
\label{eq:projector}
\end{equation}
On equal shells the number of nonzero symbols changed to zero equals the number of zeros changed to nonzero. Each is $t_h=(j-h)/2$, and $c_h=i-(j+h)/2$. The number of word pairs at the given shift is
\[
N_h=\frac{j!B^h}{t_h!^2 h!}\binom{n-j}{c_h}A^{c_h},\qquad
\max(0,2i+j-2n)\le h\le\min(j,2i-j),\quad h\equiv j\pmod 2 .
\]
Dividing the displacement by its value at zero on this shell yields
\begin{equation}
\Cx_{i,k}(j)=\frac1{D_i}\sum_h N_h
     \left[\frac{\mathcal A_k(c_h,h)}{B^k\binom ik}\right]^2 .
\label{eq:completecoef}
\end{equation}
At $k=0$ this reduces to the normalized intersection count. For positive $k$, Eq.~\eqref{eq:completecoef} gives the positive sum used to control Fourier coefficients even at a zero of one of the amplitudes.

\section{Uniform estimates for the two Fourier quantities}\label{app:stability}
\subsection{Rank-one exponents}
Take $i/n\to a$, $j/n\to\xi$, $h/n\to\eta$. The fractions of common equal nonzero positions and of positions changed from nonzero to zero are $\gamma=a-(\xi+\eta)/2$ and $v=(\xi-\eta)/2$. Stirling's formula applied to $N_h$ gives
\begin{equation}
\mathcal L_Q(a,\xi,\eta)=\xi\log_2\xi-2v\log_2v-\eta\log_2\eta
       +\eta\log_2B+\gamma\log_2A+(1-\xi)h_2\!\left(\frac{\gamma}{1-\xi}\right).
\label{eq:pairentropy}
\end{equation}
Use $0\log 0=0$ and $\eta_-=\max(0,2a+\xi-2)$, $\eta_+=\min(\xi,2a-\xi)$. The error in $\log_2 N_h-n\mathcal L_Q(i/n,j/n,h/n)$ is $O(\log n)$ uniformly up to the endpoints. At $k=0$, summing the polynomially many terms in Eq.~\eqref{eq:completecoef} gives
\begin{equation}
\begin{aligned}
D_Q(a,\xi)&=\max_{\eta_-\le\eta\le\eta_+}\mathcal L_Q(a,\xi,\eta)-H_Q(a)\\
 &=(1-\xi)\log_2(1+Ar^2)+\xi\log_2[r(2+Br)]-2a\log_2r-H_Q(a).
\end{aligned}
\label{eq:D}
\end{equation}
Here the shell-weight variable $r>0$ is the unique solution of
\begin{equation}
a=(1-\xi)\frac{Ar^2}{1+Ar^2}+\xi\frac{1+Br}{2+Br}.
\label{eq:pairr}
\end{equation}
Indeed, weighting the two words by $r$ for each nonzero entry gives the factors $1+Ar^2$ on an unchanged coordinate and $r(2+Br)$ on a changed one. Choosing $r$ so that the mean weight is $a$ gives Eq.~\eqref{eq:pairr}. Equivalently, substituting $\gamma=(1-\xi)Ar^2/(1+Ar^2)$ and $\eta=\xi Br/(2+Br)$ maximizes the strictly concave entropy in Eq.~\eqref{eq:D}. This type is interior for $0<\xi<\min(1,2a)$.

On the character side $\Ex_{i,0}(j)=|\varphi_i^{(Q,n)}(j)|^2$. The saddle equation for the coefficient in Eq.~\eqref{eq:spherical} and the squared exponent are
\begin{align}
A(1-a)z^2-[A-Ba-Q\xi]z+a&=0,\label{eq:zquad}\\
E_Q(a,\xi)&=2\big[(1-\xi)\log_2(1+Az)+\xi\log_2(1-z)-a\log_2z-H_Q(a)\big].
\label{eq:E}
\end{align}
Use the smaller positive root for $0\le\xi<g_Q(a)$. On the Cauchy circle of this radius the log modulus is concave in $\cos\theta$. Its derivative at $\cos\theta=1$ is proportional to $z[A(1-\xi)/(1+Az)^2-\xi/(1-z)^2]>0$. The real saddle is therefore the unique maximum, with positive variance. Its Gaussian contribution has positive sign and polynomial prefactor, uniformly on compact subsets up to $\xi=0$. At the turning point the roots coalesce. For estimates that cross it, character orthogonality gives $|\varphi_i(j)|^2\le Q^n/(D_iD_j)$, and averaging gives $|\varphi_i(j)|\le1$. Thus a continuous upper exponent is
\begin{equation}
U_Q^0(a,\xi)=\min\{0,\log_2 Q-H_Q(a)-H_Q(\xi)\},\qquad
U_Q(a,\xi)=
\begin{cases}
\min\{U_Q^0(a,\xi),E_Q(a,\xi)\},&\xi\le g_Q(a),\\
U_Q^0(a,\xi),&\xi>g_Q(a).
\end{cases}
\label{eq:U}
\end{equation}
Substitution of the double root $z=\sqrt{a/[A(1-a)]}$ in Eq.~\eqref{eq:E} verifies agreement at the turning point. Set $U_Q(0,\xi)=0$ where the character is identically one.

\subsection{From one shell to the annulus}
Convexity in Eq.~\eqref{eq:mixture} gives $\Psi(\vec{x})\le\max_i\Ex_{i,k}(j)$. A character block specified by the input and output anchor sets has a color matrix independent of the shell, multiplied by the radial amplitude in Eq.~\eqref{eq:migration}. In a region of positive real saddles those amplitudes have the same sign across the annulus. Retaining such a block in shell $i$ gives a lower bound with factor $p_i^2$. Some $p_i$ is at least $1/|I_n|$, which has zero cost per coordinate.

Every mixed-shell displacement in Eq.~\eqref{eq:master-C} is nonnegative, so the same shell choice also transfers the lower bound on $\Cx_{i,k}$. Explicitly, the diagonal-shell contribution to $\nu(\vec{x})/\nu(\vec{0})$ has coefficient $(p_i^2/D_i)/\sum_\ell(p_\ell^2/D_\ell)$ multiplying $\Cx_{i,k}$. For such a shell that coefficient is $2^{o(n)}$, since all shell sizes have the same exponent. To bound the mixed terms from above, Eq.~\eqref{eq:projector} gives
\[
|\mathcal A_k(c,h)|\le B^k\binom{c+h}k
 \le B^k\sqrt{\binom ik\binom\ell k},\qquad
|P_{i,\ell;k}(\vec{u},\vec{v})|\le\frac{m_k}{\sqrt{D_iD_\ell}} .
\]
In their pair counts, $t_h,t_h,c_h$ are replaced by $(j-h+i-\ell)/2,(j-h-i+\ell)/2,(i+\ell-j-h)/2$. Since both radii tend uniformly to $a$, the limiting count exponent remains the maximum in Eq.~\eqref{eq:D} before subtracting $H_Q(a)$. Normalizing by $\nu(\vec{0})$ and summing over shells gives $d_n(\xi)\le D_Q(a,\xi)+o_n(1)$.

\subsection{The estimates near and away from zero}
We first estimate the displacement from below. Put $N=c+h$. The amplitude $\mathcal A_k(c,h)$ equals the degree-$k$ Krawtchouk polynomial for $B+1$ symbols evaluated at $h$~\cite{MacWilliamsSloane}. Its three-term recurrence for multiplication by $h$ in the orthonormal basis has diagonal $[BN-(B-1)\ell]/(B+1)$ and adjacent coefficient $\sqrt{B(\ell+1)(N-\ell)}/(B+1)$ in absolute value. The zeros are eigenvalues of the leading $k$-dimensional tridiagonal matrix. The row bounds place them above $\zeta_{N,k}=[BN-(B-1)(k-1)-2\sqrt{BkN}]/(B+1)$. Factoring the polynomial by these zeros and using its value $B^k\binom Nk$ at zero gives, for $h<\zeta_{N,k}$,
\begin{equation}
\frac{\mathcal A_k(c,h)}{B^k\binom ik}
\ge\left[\frac{N-k+1}{i-k+1}\left(1-\frac{h}{\zeta_{N,k}}\right)\right]^k .
\label{eq:root-bound}
\end{equation}
If $k\le i/128$ and $j\le i/16$, then $N\ge31i/32$ and $\zeta_{N,k}\ge N/3$. The right side is at least $\exp(-8kj/i)$. Squaring in Eq.~\eqref{eq:completecoef} gives $d_n\ge D_Q-Cb\xi-o_n(1)$, including bounded $j$. At a maximizing positive-weight type with $B\gamma>\eta$, the limiting separation of $h/n$ from $\zeta_{N,k}/n$ is positive for small $b$. Keeping that one type gives $d_n\ge D_Q-Cb-o_n(1)$ uniformly in a compact neighborhood.

At other positive weights an amplitude at the maximizing type could vanish. To control the sum, keep $k+1$ consecutive physical types $h,h+2,\ldots,h+2k$, and denote their weights by $W_\ell=N_{h+2\ell}$. The following finite difference fixes one linear combination of their amplitudes:
\begin{align}
\sum_{\ell=0}^k(-1)^\ell\binom k\ell\mathcal A_k(c-\ell,h+2\ell)
 &=[z^k](1+Bz)^{c-k}(1-z)^h\big[(1+Bz)-(1-z)^2\big]^k=Q^k,\nonumber\\
\sum_{\ell=0}^k W_\ell|\mathcal A_k(c-\ell,h+2\ell)|^2
 &\ge\frac{Q^{2k}}{\displaystyle\sum_{\ell=0}^k\binom k\ell^2/W_\ell}.
\label{eq:weightedblock}
\end{align}
The second line is weighted Cauchy--Schwarz. On a compact positive-weight interval strictly inside $\xi<\min(1,2a)$, the maximizing type is interior. Starting a block there changes each pair-count exponent by at most $Cb$. The denominator in Eq.~\eqref{eq:weightedblock} is at most $4^k/\min W_\ell$. The normalization in Eq.~\eqref{eq:completecoef} then gives $d_n\ge D_Q-2h_2(b)-Cb-o_n(1)$.

For the character near zero, put $\theta=t/n$, and let $p=(a-b-\theta)/(1-b-\theta)$ and $x=(\xi-r/n-\theta)/(1-b-\theta)$ be the shifted radius and weight in Eq.~\eqref{eq:migration}. Then $|p-a|\le Cb$, $|x-\xi|\le C(b\xi+r/n)$, and Eq.~\eqref{eq:E} gives $E_Q(a,\xi)=O(\xi)$ and $\partial_aE_Q(a,\xi)=O(\xi)$. It follows that
\[
(1-b-\theta)E_Q(p,x)\le E_Q(a,\xi)+Cb\xi+C'r/n .
\]
The saddle bound is uniform up to a polynomial prefactor. The remaining conditioning factor and the local squared weights are at most one. The exchange binomials sum to $\binom jr$ at fixed $r$. If $R$ is the hypergeometric intersection size, then $\Pr(R=r)\le\binom jr(k/n)^r$, by fixing any $r$ prescribed positions. Consequently, for $v>0$,
\[
\mathbb E\!\left[\binom jR v^R\right]\le
 \sum_r\binom jr^2(kv/n)^r\le(1+\sqrt{kv/n})^{2j}.
\]
Taking $v=2^{C'}$ gives the character estimate in Eq.~\eqref{eq:local-bounds}. Away from zero, $\binom rt\le2^k$ and $\binom{j-r}t\le\binom nt\le2^{nh_2(k/n)}$ for small $k/n$. Replacing the shifted character exponent by its upper envelope in Eq.~\eqref{eq:U} costs at most a modulus of continuity $\omega(b)\to0$. Together with the displacement estimate this proves the uniform bounds
\begin{equation}
e_n\le U_Q(a,\xi)+h_2(b)+Cb+\omega(b)+o_n(1),\qquad
d_n\ge D_Q(a,\xi)-2h_2(b)-Cb-o_n(1).
\label{eq:bulk-bounds}
\end{equation}
Before the turning point the exponent is smooth and the replacement costs $O(b)$. The matching lower bound in Eq.~\eqref{eq:interior-bounds} comes from the term $r=t=k$ of Eq.~\eqref{eq:migration},
\[
\frac{\binom jk}{\binom nk}\binom{j-k}k\,\beta_Q^k
 \left[\frac{\binom{n-2k}{i-2k}}{\binom{n-k}{i-k}}\,
       \varphi_{i-2k}^{(Q,n-2k)}(j-2k)\right]^2 .
\]
At fixed interior positive weight $\binom{j-k}k$ contributes $h_2(b)+O(b)$ per coordinate. The remaining factors shift the rank-one exponent by $O(b)$. The radial saddle is positive across the shells for this term, so the lower estimate also holds for the annulus.

\subsection{Using all physical types}
Equation~\eqref{eq:weightedblock} uses only part of the positive energy. On the full physical lattice $h_\ell=h_0+2\ell$, $0\le\ell\le M$, write $p(\ell)=\mathcal A_k(c_0-\ell,h_0+2\ell)$ and $W_\ell=N_{h_\ell}$. The generating polynomial is $(1+Bz)^{c_0}(1-z)^{h_0}\exp\{\ell[2\ln(1-z)-\ln(1+Bz)]\}$. The bracket starts with $-Qz$, so $p$ has degree $k$ and leading coefficient $L_k=(-Q)^k/k!$. Write $\vec{v}$ for its monomial coefficient vector. If $M\ge k$, the moment matrix $\mathbf G_{rs}=\sum_\ell W_\ell\ell^{r+s}$, $0\le r,s\le k$, is positive definite. Fix any collection of coefficients by $S\vec{v}=\vec{d}$, with $S$ selecting distinct entries. Minimizing the energy subject to these data gives
\begin{equation}
\sum_\ell W_\ell|p(\ell)|^2=\vec{v}^\dagger\mathbf G\vec{v}
 \ge \vec{d}^\dagger(S\mathbf G^{-1}S^\dagger)^{-1}\vec{d} .
\label{eq:projection}
\end{equation}
Indeed the minimum is attained by $\mathbf G^{-1}S^\dagger(S\mathbf G^{-1}S^\dagger)^{-1}\vec{d}$. The leading coefficient alone gives $L_k^2/(\mathbf G^{-1})_{kk}$. Every polynomial with that coefficient satisfies the finite difference in Eq.~\eqref{eq:weightedblock}, so the full-lattice bound is at least as strong as any one block bound. Additional coefficients can be imposed in the same minimization.

\section{Exchange of anchor positions}\label{app:exchange}
\subsection{The weighted selection problem}
In Eq.~\eqref{eq:reservoir} the ratio of the $t$th summand to $\binom kt(kx^2)^t/t!$ is $\prod_{r=0}^{t-1}(1-r/M)$. Since $-z/(1-z)\le\ln(1-z)\le0$, the finite sum satisfies
\begin{equation}
e^{-k(k-1)/[2(M-k+1)]}\,\mathrm L_k(-kx^2)\le Z_{k,M}(x)\le\mathrm L_k(-kx^2),\qquad
\mathrm L_k(-kx^2)=\sum_{t=0}^k\binom kt\frac{(kx^2)^t}{t!}.
\label{eq:laguerre}
\end{equation}
Here $\mathrm L_k$ is the ordinary Laguerre polynomial~\cite{DLMF}. When $k/M\to0$, the logarithms per $k$ of the three quantities have the same limit. At $t/k=v$, Stirling's formula gives the exponent of the Laguerre summand and its maximum as
\begin{equation}
\begin{aligned}
\lim_{k\to\infty}\frac1k\log_2 Z_{k,M}(x)
 &=\max_{0\le v\le1}\{h_2(v)-v\log_2v+v\log_2 e+2v\log_2x\}\\
 &=\frac{v(x)-\ln[1-v(x)]}{\ln2},\qquad
v(x)=\frac{2x}{\sqrt{x^2+4}+x}.
\end{aligned}
\label{eq:pressure}
\end{equation}
The first line is strictly concave and its stationary equation is $v^2=x^2(1-v)$. The unique maximum proves concentration of $t/k$ at $v(x)$ and convergence uniform on compact positive $x$-intervals. Differentiating at the maximum gives $d\mathfrak F/d\log_2 x=2v(x)$.

\subsection{Reduction of the Hamming sum}
Let $\alpha=r/n,\ \theta=t/n$ in Eq.~\eqref{eq:migration}. The exponent with squared radial character exponent $V$ is
\begin{equation}
\begin{aligned}
\mathfrak M_V(a,b,\xi;\alpha,\theta)
={}&\xi h_2(\alpha/\xi)+(1-\xi)h_2\!\left(\frac{b-\alpha}{1-\xi}\right)-h_2(b)
      +\alpha h_2(\theta/\alpha)+(\xi-\alpha)h_2\!\left(\frac{\theta}{\xi-\alpha}\right)\\
 &+(\alpha-\theta)\log_2\alpha_Q+\theta\log_2\beta_Q
      +2(1-b-\theta)h_2(p)-2(1-b)h_2\!\left(\frac{a-b}{1-b}\right)+(1-b-\theta)V(p,x),
\end{aligned}
\label{eq:M}
\end{equation}
where
\begin{equation}
\begin{gathered}
p=\frac{a-b-\theta}{1-b-\theta},\qquad x=\frac{\xi-\alpha-\theta}{1-b-\theta},\\
\max(0,b+\xi-1)\le\alpha\le\min(b,\xi),\qquad
0\le\theta\le\min(\alpha,\xi-\alpha,a-b).
\end{gathered}
\label{eq:Mdomain}
\end{equation}
All entropy terms take their continuous values on the boundary. There are only polynomially many integer pairs $(r,t)$, so the supremum with $V=U_Q$ bounds $e_n$ in the length limit. Where all the shifted radial saddles are stable, the character amplitudes in each block have common sign across the shells. The supremum with $V=E_Q$ then gives the exact exponent.

Fix $a$ in a compact subset of $(0,A/Q)$ and put $\xi=u\varepsilon$, $\varepsilon=\sqrt b$, where $u$ ranges over a compact positive interval. Set $\alpha=b\sigma$, $\theta=b\tau$. For small $b$ the domain is $0\le\tau\le\sigma\le1$. With $\mu_0=1-Qa/A$, Eq.~\eqref{eq:M} gives
\begin{equation}
\begin{aligned}
\frac{\mathfrak M_{E_Q}-E_Q(a,u\varepsilon)}{b}
={}&(\sigma-\tau)\log_2\varepsilon+h_2(\sigma)+\sigma h_2(\tau/\sigma)
  +(\sigma+\tau)\log_2u-\tau\log_2\tau+\tau\log_2 e\\
 &+(\sigma-\tau)\log_2\alpha_Q+\tau\log_2\beta_Q
       +2\tau\log_2 a-2(\sigma+\tau)\log_2\mu_0+o(1).
\end{aligned}
\label{eq:crossover-expansion}
\end{equation}
The hypergeometric probability contributes $\sigma\log_2\varepsilon+\sigma\log_2u+h_2(\sigma)$. Choosing gained positions contributes $-\tau\log_2\varepsilon+\tau\log_2u-\tau\log_2\tau+\tau\log_2e$. The conditioning ratio contributes $2\tau\log_2a$. The shift of the radial character follows from $E_Q(a,0)=\partial_aE_Q(a,0)=0$ and $\partial_\xi E_Q(a,0)=2\log_2\mu_0$. These expansions have uniform remainder on the triangle by smoothness at the radial saddle and continuity of $t\log t$ at zero.

The term $(\sigma-\tau)\log_2\varepsilon$ forces every maximizing sequence to approach the diagonal. There the expression in Eq.~\eqref{eq:crossover-expansion} is precisely the objective in Eq.~\eqref{eq:pressure} at $v=\sigma=\tau$, $x=c_a u$. Thus $r/k$ and $t/k$ tend to $v(c_a u)$ in the successive limits $n\to\infty$, $b\downarrow0$. Equivalently, the difference between $\log_2\Psi(\vec{x})-nE_Q(a,u\sqrt b)$ and $\log_2Z_{k,j}(c_a u)$, divided by $bn$, tends to zero in these limits for $j=\wt(\vec{x})$ and $j/n\to u\sqrt b$. Equation~\eqref{eq:local-bounds} and the mixed-shell upper bound put $d_n$ between $D_Q-Cb\xi-o_n(1)$ and $D_Q+o_n(1)$, giving the stated $o(b)$ displacement correction. At small $x$, $\mathfrak F(x)=2x/\ln2+O(x^2)$, whereas at large $x$, $\mathfrak F(x)=2\log_2x+\log_2e+O(x^{-2})$. Equation~\eqref{eq:interior-bounds} separately treats a fixed positive $\xi$.

\subsection{Fixed error weight and the slope at zero}
The small-weight end admits a second calculation that keeps $j$ fixed as $n\to\infty$. Define the return polynomial
\[
P_j(x,y)=[z^0]\big(x+\sqrt y\,(z+z^{-1})\big)^j
 =\sum_{t=0}^{\lfloor j/2\rfloor}\frac{j!}{t!^2(j-2t)!}\,x^{j-2t}y^t .
\]
For $0<b<Ba/A$, put $p=(a-b)/(1-b)$, $\mu=1-Qp/A$, and define
\begin{equation}
\begin{aligned}
e&=b\alpha_Q+(1-b)\mu^2,& w&=b(1-b)\beta_Qp^2,\\
d_0&=\frac{Ba}{A}\left(1-\frac{Ab}{Ba}\right)^2,&
u_0&=\frac{a(1-a)}{A}\left(1-\frac ba\right)^2 .
\end{aligned}
\label{eq:walk-weights}
\end{equation}
In Eq.~\eqref{eq:migration} the finite-position limits of the intersection probability, the conditioning ratio and the radial character are $\binom jr b^r(1-b)^{j-r}$, $p^t$, and $\mu^{j-r-t}$, respectively. Combining the $r-t$ retained anchors with the $j-r-t$ unchanged radial positions gives
$\Ex_{i,k}(j)\to P_j(e,w)$. In Eq.~\eqref{eq:completecoef}, $t=(j-h)/2$ and
\[
\frac{N_h}{D_i}\longrightarrow\frac{j!}{t!^2h!}
  \left(\frac{a(1-a)}A\right)^t\left(\frac{Ba}A\right)^h,\qquad
\frac{\mathcal A_k(c_h,h)}{B^k\binom ik}\longrightarrow
  (1-b/a)^t\left(1-\frac{Ab}{Ba}\right)^h.
\]
This yields $\Cx_{i,k}(j)\to P_j(d_0,u_0)$. The growth factors of these polynomials as $j\to\infty$ are $\lambda_E=e+2\sqrt w$ and $\lambda_C=d_0+2\sqrt{u_0}$, respectively, by maximizing the multinomial entropy with equal numbers of the two opposite steps.

To obtain the slope with the length limit taken first, expand Eq.~\eqref{eq:M} at fixed $b$ and small $\xi$. Set $\alpha=s'\xi$, $\theta=t'\xi$ and write $(f_1,f_2,f_3,f_4)=(s'-t',1-s'-t',t',t')$. The first-order term is
\[
\xi\left[-\sum_i f_i\log_2 f_i+f_1\log_2(b\alpha_Q)
 +f_2\log_2((1-b)\mu^2)+t'\log_2 w\right].
\]
The remainder is $O_b(\xi^2)$ uniformly on the rescaled simplex. The maximum of the bracket is $\log_2\lambda_E$. For the displacement, at $\xi=0$ the coefficient of $(1+Bz)^{an}$ has its positive saddle at $y_0=b/[B(a-b)]<1$. For $\eta=\ell\xi$, $t'=(1-\ell)/2$, expanding the logarithm of $(1+By)^{n\gamma}(1-y)^{n\eta}y^{-nb}$ at its stationary $y$ gives the normalized amplitude term $\xi[t'\log_2(1-b/a)+\ell\log_2(1-Ab/(Ba))]$. After squaring and adding Eq.~\eqref{eq:pairentropy}, the three multinomial weights are $d_0,\sqrt{u_0},\sqrt{u_0}$. Their maximum exponent is $\log_2\lambda_C$. The positive saddles persist uniformly for small $\xi$ and for the shell radii in the annulus. The mixed-shell pair counts converge to the same entropy, and retaining one shell gives the matching lower exponent. Hence
\begin{align}
\lim_{n\to\infty}(e_n-d_n)&=\xi\log_2(\lambda_E/\lambda_C)+O_b(\xi^2),\label{eq:low-slope}\\
\log_2(\lambda_E/\lambda_C)&=\partial_\xi\Gamma_Q(a,0)
  +\frac{2a\sqrt{\beta_Q}}{(1-Qa/A)^2\ln2}\sqrt b+O(b).\label{eq:sqrt-sharp}
\end{align}
The second line follows by expanding Eq.~\eqref{eq:walk-weights} at $b=0$. Its positive square-root term agrees with the small-$x$ limit of Eq.~\eqref{eq:crossover}.

\section{The first contacts and their continuation}\label{app:geometry}
\subsection{Sign of the unperturbed gap}
Differentiating the two optimized exponents in Eqs.~\eqref{eq:D} and \eqref{eq:E} at their saddles gives, with $z_0=a/[A(1-a)]$,
\begin{equation}
\partial_\xi E_Q=2\log_2\frac{1-z}{1+Az},\qquad
\partial_\xi D_Q=\log_2\frac{r(2+Br)}{1+Ar^2},\qquad
\partial_a\Gamma_Q=\log_2\frac{r^2z_0}{z^2}.
\label{eq:gapderiv}
\end{equation}
At zero weight $r=\sqrt{z_0}$ and $z=z_0$. As $\xi$ increases, Eq.~\eqref{eq:pairr} makes $r$ decrease and Eq.~\eqref{eq:zquad} makes the smaller root $z$ increase. Thus $\partial_a\Gamma_Q<0$ on $0<\xi<\min(2a,g_Q(a))$. This monotonicity will allow a sign at one radius to control all larger radii on their physical weight intervals.

A stationary value in $\xi$ must satisfy
\begin{equation}
S_A(r,z):=r(2+Br)(1+Az)^2-(1+Ar^2)(1-z)^2=0.
\label{eq:Epoly}
\end{equation}
To recover $a,\xi$ at that point, write the two saddle equations in the form $a=L(z)-\xi C(z)=M(r)+\xi D(r)$, where
\[
L(z)=\frac{Az}{1+Az},\quad C(z)=\frac{Qz}{(1+Az)(1-z)},\quad
M(r)=\frac{Ar^2}{1+Ar^2},\quad D(r)=\frac{(1-r)(1+Ar)}{(2+Br)(1+Ar^2)} .
\]
Solving this pair of linear equations gives
\begin{equation}
\xi=\frac{L(z)-M(r)}{C(z)+D(r)},\qquad
a=\frac{L(z)D(r)+M(r)C(z)}{C(z)+D(r)} .
\label{eq:stationaryvalues}
\end{equation}
The polynomial $S_A(r,z)$ is symmetric in $r,z$. With $w=\sqrt{2/(rz)+A(A-1)}-A$, its positive curve can be written as
\begin{equation}
rz=\frac{2}{w^2+2Aw+A},\qquad r+z=\frac{w^2-2w-A}{w^2+2Aw+A}.
\label{eq:wparam}
\end{equation}
Indeed Eq.~\eqref{eq:Epoly} gives $z=(1-y)/(1+Ay)$, $y=\sqrt{r(2+Br)/(1+Ar^2)}$. This branch is connected for $0<r<1$, and $w\to\infty$ at $r\to0$. Since $w=0$ would give $r+z=-1$, positivity of the sum requires $w>q+1$. The numerator of $(r+z)^2-4rz$ is
\[
[w^2-2(q+1)w+q^2-2q-3]\,[w^2+2(q-1)w+q^2+2q-3].
\]
The two real positive roots thus require $w\ge w_c=q+1+2\sqrt{q+1}$. Substituting Eq.~\eqref{eq:wparam} into Eq.~\eqref{eq:stationaryvalues} gives
\begin{equation}
a=T_A(w)=\frac{A(A+w^2-2)}{w^3+3Aw^2+Aw+A(A-2)} .
\label{eq:Tw}
\end{equation}
The derivative has numerator $-A[w^4+(2A-6)w^2+4A(A-2)w+A(A-2)]$ over the square of the denominator, so $T_A$ is strictly decreasing. At $w_c$, $T_A(w_c)=\vartheta_q$ in Eq.~\eqref{eq:threshold}. Consequently there are no stationary points for $a>\vartheta_q$. For $0<a<\vartheta_q$ there is one root $w>w_c$ of
\begin{equation}
aw^3+A(3a-1)w^2+Aaw+A(A-2)(a-1)=0 .
\label{eq:cubic}
\end{equation}
The two orderings of the roots in Eq.~\eqref{eq:wparam} give all candidates. An interior stationary point must also have $0<\xi<g_Q(a)$ and $z<\sqrt{a/[A(1-a)]}$.

Near $a=A/Q$ the origin slope is negative, or equivalently $S_A(\sqrt{z_0},z_0)>0$. For $a>\vartheta_q$ this sign cannot change at zero or at another physical weight, by the bound on stationary radii. The pair saddle persists up to $\xi=g_Q(a)$, since otherwise $r\to0$ at $2a\le g_Q(a)$ would give $S_A<0$. Hence $\Gamma_Q(a,\xi)<0$ for $0<\xi\le g_Q(a)$. Compactness gives a uniform bound $-c\xi$ near zero and a fixed negative bound beyond that neighborhood. This proves the interval in Theorem~\ref{thm:lossless} for every $q$. For the three smallest dimensions the two stationary candidates allow a larger interval to be determined.

\subsection{Location of the first contact}
At a stationary point the coefficient of $\xi$ in $E_Q-D_Q$ vanishes by Eq.~\eqref{eq:Epoly}. The remaining expression is
\begin{equation}
\Gamma_Q(a,\xi)=2\log_2(1+Az)-\log_2(1+Ar^2)+2a\log_2(r/z)-H_Q(a).
\label{eq:stationarygap}
\end{equation}
Let $\xi_+(a),\xi_-(a)$ denote the weights corresponding to $r<z$ and $r>z$, respectively. For each $q=2,3,4$, take $a^-=N_q/10^{10}$ and $a^+=(N_q+1)/10^{10}$, with
\[
(N_2,N_3,N_4)=(1899656480,\ 2562028752,\ 2909720938).
\]
The root of Eq.~\eqref{eq:cubic} lies in $(13/2,7)$, $(33/4,17/2)$, and $(19/2,10)$, respectively. At each endpoint $a^\pm$, bisection on this interval isolates $w$. Since $T_A$ is decreasing, the endpoint brackets also enclose $w$ for every intermediate $a$. Substitution in Eqs.~\eqref{eq:wparam} and \eqref{eq:stationaryvalues} puts both stationary candidates strictly inside the physical interval on the smaller character-root branch.

\begin{table}[tb]
\caption{Sign estimates used to isolate the first positive-weight contact. The first two bounds are evaluated at the indicated radius. The remaining bounds hold throughout $[a^-,a^+]$ for each of $q=2,3,4$. The factor $\ln 2$ converts a gap in bits to natural-log units.}
\label{tab:signs}
\begin{ruledtabular}
\begin{tabular}{lc}
Quantity & Bound\\
\hline
$(\ln2)\,\Gamma_Q(a^-,\xi_+(a^-))$ & $>7\times10^{-11}$\\
$(\ln2)\,\Gamma_Q(a^+,\xi_+(a^+))$ & $<-6\times10^{-11}$\\
$(\ln2)\,\Gamma_Q(a,\xi_-(a))$ & $<-3\times10^{-3}$\\
$(\ln2)\,\Gamma_Q(a,g_Q(a))$ & $<-9\times10^{-4}$\\
$S_A(\sqrt{z_0},z_0)$ & $>1/50$\\
$B\gamma-\eta$ at $\xi_+(a)$ & $>19/1000$\\
$-\partial_a\Gamma_Q(a,\xi_+(a))-H'_Q(a)$ & $>9/5$
\end{tabular}
\end{ruledtabular}
\end{table}

The estimates in Table~\ref{tab:signs} can be obtained with rational endpoints. Bisect the cubic 150 times, enclosing square roots by squaring their endpoints. For the turning-point gap take the double root for $z$ and bisect the increasing right side of Eq.~\eqref{eq:pairr} for $r$. To enclose a logarithm, scale its argument by a power of two to $1\le y\le2$ and put $t=(y-1)/(y+1)$. With $N=64$,
\[
0\le\ln y-2\sum_{i=0}^{N-1}\frac{t^{2i+1}}{2i+1}
 \le\frac{2t^{2N+1}}{(2N+1)(1-t^2)} .
\]
Taking $y=2$ also encloses $\ln2$. These computations give opposite signs at the endpoints for the $r<z$ branch. Its gap decreases strictly in $a$ by Eq.~\eqref{eq:gapderiv}, so it vanishes at a unique $a_c$ in the bracket. At this radius the other stationary gap and the turning-endpoint gap are negative and the slope at zero is negative. Thus the only positive-weight zero is $\xi_c=\xi_+(a_c)$. Monotonicity in $a$ gives a strict gap at every larger radius up to its own turning endpoint. This proves that the contact is the first encountered as the relative distance increases, with $g_Q(a^+)<\delta_c<g_Q(a^-)$. It also establishes the uniform signs used in the proof of Theorem~\ref{thm:lossless}.

At the contact Eq.~\eqref{eq:gapderiv} gives $G=H'_Q(a_c)+2\log_2(z/r)$. The branch $r<z$ therefore has $G>H_a$. For qubits the same endpoint calculation gives
\[
\frac{181}{1000}G>1+10^{-5},\qquad
1-H_a\left(\frac{181}{1000}+\frac{1}{10|g'_4(a_c)|}\right)>\frac{23}{1000}.
\]
Choosing $\tau=181/1000$ and $\sigma=1/10$ in Eq.~\eqref{eq:contact-rate} removes the positive-weight excess and leaves a gain exceeding $h_b/50$ for sufficiently small $b$. This proves Eq.~\eqref{eq:qubit}.

\subsection{A compact set of contacts}\label{app:directions}
For a spectral family with parameters $\vec{\theta}$, let $\ell_b(\vec{\theta},\xi)$ be the length-limit logarithmic ratio per coordinate of $\nu(\vec{0})\Psi(\vec{x})/\nu(\vec{x})$, and $h_b(\vec{\theta})$ the corresponding origin entropy per coordinate. In this subsection write $\epsilon_b\downarrow0$ for the scale of the gain. Suppose the continuous unperturbed gap $\Gamma(\vec{\theta}_c,\xi)$ on a compact error set is nonpositive and vanishes exactly at the origin and a compact set $\mathcal Z$ separated from it. For a fixed direction $\vec{v}$ realized by admissible perturbations $\vec{\theta}_b=\vec{\theta}_c+\epsilon_b\vec{v}+o(\epsilon_b)$, assume that the maximum in a neighborhood of zero is zero, and that away from this neighborhood and those of $\mathcal Z$ there is a uniform strict gap. Near $\mathcal Z$ assume uniform expansions
\begin{equation}
\begin{aligned}
\ell_b(\vec{\theta}_b,\xi)&=\Gamma(\vec{\theta}_c,\xi)+
 \epsilon_b[\alpha(\xi)-\vec{g}_\xi\cdot\vec{v}]+o(\epsilon_b),\qquad
 \vec{g}_\xi=-\nabla_{\vec{\theta}}\Gamma(\vec{\theta}_c,\xi),\\
h_b(\vec{\theta}_b)&=h_0(\vec{\theta}_c)+\epsilon_b(\vec{h}\cdot\vec{v}-\beta)+o(\epsilon_b),
\end{aligned}
\label{eq:general-expansion}
\end{equation}
with continuous contact data. These assumptions can also be used with uniform upper and lower length exponents.

Every convergent sequence of maximizing errors outside the origin neighborhood must approach $\mathcal Z$. The first line of Eq.~\eqref{eq:general-expansion} bounds its first-order excess from above by the largest value of $\alpha(\xi)-\vec{g}_\xi\cdot\vec{v}$ on $\mathcal Z$. Evaluating at a contact attaining that value gives the lower bound, and evaluating at the origin gives zero. Hence
\begin{equation}
\sup_\xi\ell_b(\vec{\theta}_b,\xi)=
 \epsilon_b\max\left\{0,\max_{\xi\in\mathcal Z}[\alpha(\xi)-\vec{g}_\xi\cdot\vec{v}]\right\}
 +o(\epsilon_b).
\label{eq:general-contact}
\end{equation}
If $\vec{g}_\xi\cdot\vec{v}>\alpha(\xi)$ throughout $\mathcal Z$, compactness gives a strict margin there and the length-limit excess is zero for all sufficiently small $b$. For the annular dual, Eq.~\eqref{eq:interior-bounds} supplies $\alpha=\beta=1$ and the adjacent-weight estimate in Sec.~\ref{sec:contact} identifies the same first-order excess for $f/\widehat f$.

For $\alpha=\beta=1$ let the admissible directions form a closed convex cone $\mathcal V$. On a ray $t\vec{v}$, $t\ge0$, Eq.~\eqref{eq:cost} gives a negative cost at some $t$ exactly when $\inf_{\xi\in\mathcal Z}(\vec{g}_\xi-\vec{h})\cdot\vec{v}>0$. Indeed, for small positive $t$ the cost is $t(\vec{h}\cdot\vec{v}-\inf_\xi\vec{g}_\xi\cdot\vec{v})$, and this is a lower bound on the cost at any $t$. Writing $\mathcal V^*=\{\vec{y}:\vec{y}\cdot\vec{v}\ge0\text{ for every }\vec{v}\in\mathcal V\}$, strict separation of a compact convex set and a closed cone gives the equivalent criterion~\cite{BoydVandenberghe}
\[
\operatorname{conv}\{\vec{g}_\xi:\xi\in\mathcal Z\}\cap(\vec{h}-\mathcal V^*)=\varnothing .
\]
When $\vec{h}\in\mathcal V^*$, scaling each improving direction until its least suppression rate is one gives the optimal retained fraction $[1-c_*]_+$, where
\begin{equation}
c_*=\inf\{\vec{h}\cdot\vec{v}:\ \vec{v}\in\mathcal V,\ \vec{g}_\xi\cdot\vec{v}\ge1\text{ for every }\xi\in\mathcal Z\}.
\label{eq:covering}
\end{equation}
An empty feasible set gives zero retained fraction. With finitely many contacts and a polyhedral $\mathcal V$, the dual of the feasible minimization maximizes $\sum_\ell y_\ell$ over $y_\ell\ge0$ with $\vec{h}-\sum_\ell y_\ell\vec{g}_{\xi_\ell}\in\mathcal V^*$. In the one-radius case Eq.~\eqref{eq:covering} is simply $c_*=H_a/G$.

\section{Rate bound at finite anchor density}\label{app:finite}
Fix a radius and density satisfying
\begin{equation}
0<a<A/Q,\qquad 0\le b<\frac BA(a-\delta/2),\qquad
a+\delta/2<1,\qquad \Lambda_Q(a,b)>1-Q\delta/A .
\label{eq:admissible}
\end{equation}
We use Eq.~\eqref{eq:transfer} without expanding at $b=0$. Equation~\eqref{eq:M} gives the upper exponent $\mathcal E_Q(a,b,\xi)=\sup_{\alpha,\theta}\mathfrak M_{U_Q}(a,b,\xi;\alpha,\theta)$ on the domain in Eq.~\eqref{eq:Mdomain}. The following three lower estimates for $d_n$ come directly from the sum of squares in Eq.~\eqref{eq:completecoef}.

First keep the term $h=0$ for even $j$. Its amplitude is $B^k\binom{i-j/2}k$. Inserting the pair count $N_0$ and applying Stirling's formula gives the exponent
\begin{equation}
C_{\rm sparse}=\xi-h_2(a)-\frac{\xi}{2}\log_2 A
 +(1-\xi)h_2\!\left(\frac{a-\xi/2}{1-\xi}\right)
 +2(a-\xi/2)h_2\!\left(\frac{b}{a-\xi/2}\right)-2a h_2(b/a).
\label{eq:sparse}
\end{equation}
For odd $j$, take $h=1$. Then $\mathcal A_k(c,1)=B^k\binom ck[1-k/(B(c-k+1))]$ and the bracket stays positive under Eq.~\eqref{eq:admissible}. This gives the same lower exponent and proves positivity at every required weight for all sufficiently large $n$.

Next retain any interior physical type $\eta$ with a stable real saddle for the amplitude, writing $\gamma=a-(\xi+\eta)/2$. For $b>0$ the saddle $0<y<1$ must satisfy
\begin{equation}
b=\frac{B\gamma y}{1+By}-\frac{\eta y}{1-y},\qquad
\frac{B\gamma}{(1+By)^2}-\frac{\eta}{(1-y)^2}>0 .
\label{eq:color-saddle}
\end{equation}
On $z=ye^{i\theta}$ the log modulus of $(1+Bz)^{n\gamma}(1-z)^{n\eta}$ relative to its value at $y$ is at most $-ny[B\gamma/(1+By)^2-\eta/(1-y)^2](1-\cos\theta)$. The Gaussian contribution at the saddle has positive sign and the part outside any fixed neighborhood is exponentially smaller. The exact exponent of the amplitude is thus $J_b(\gamma,\eta)=\gamma\log_2(1+By)+\eta\log_2(1-y)-b\log_2y$. Including its pair count gives
\begin{equation}
C_{\rm stable}=\sup_{\eta,y}
 \big\{\mathcal L_Q(a,\xi,\eta)-H_Q(a)+2[J_b(\gamma,\eta)-b\log_2 B-a h_2(b/a)]\big\},
\label{eq:stable}
\end{equation}
where the supremum uses Eq.~\eqref{eq:color-saddle} and interior types, with endpoint values taken by interior limits.

Finally, Eq.~\eqref{eq:weightedblock} gives a lower bound even when individual amplitudes change sign or vanish. For a block starting at $\eta$ the denominator is a sum with terms $\binom k\ell^2/N_{h+2\ell}$. Maximizing their exponents over $\tau=\ell/n$ and then choosing the best starting type gives
\begin{equation}
C_{\rm diff}=2b\log_2(Q/B)-2a h_2(b/a)-H_Q(a)
 -\inf_{\eta_-\le\eta\le\eta_+-2b}\ 
  \sup_{0\le\tau\le b}\{2b h_2(\tau/b)-\mathcal L_Q(a,\xi,\eta+2\tau)\}.
\label{eq:diff-exp}
\end{equation}
All three bounds refer to the same normalized displacement. Set $C_Q^*=\max\{C_{\rm sparse},C_{\rm stable},C_{\rm diff}\}$, assigning value $-\infty$ to an unavailable branch. At a weight where a branch first becomes available use the lower limiting value of this maximum over neighboring admissible weights, including the weight itself. This convention covers integer sequences approaching the boundary from either side. At $b=0$ use $C_Q^*=D_Q$ directly, and at $\xi=0$ both normalized exponents are zero. We obtain
\begin{equation}
\begin{aligned}
\Delta_Q^*(a,b;\delta)&=\sup_{0\le\xi\le\delta}
             [\mathcal E_Q(a,b,\xi)-C_Q^*(a,b,\xi)],\\
R_q(\delta)&\le
 \left[\frac{\Origin(a,b)+\Delta_Q^*(a,b;\delta)}{\log_2q}-1\right]_+ .
\end{aligned}
\label{eq:finite-bound}
\end{equation}
To pass to the weight maximum, take a maximizing sequence $j/n$ and a convergent subsequence. The upper envelope in Eq.~\eqref{eq:U} is continuous. Every strictly stable interior type persists under small changes of the weight. The finite-difference estimate persists wherever there is room for the block, and the sparse estimate is available throughout. These estimates transfer from shells to the annulus as in Appendix~\ref{app:stability}. Equation~\eqref{eq:transfer} then gives Eq.~\eqref{eq:finite-bound}, with the positive part accounting for the convention in Eq.~\eqref{eq:rate}. Taking the infimum over admissible $(a,b)$ gives the finite-density optimization.

\end{document}